\documentclass[showpacs,showkeys,amsmath,amssymb,floatfix]{revtex4}
\usepackage{amsfonts}
\usepackage{graphicx}

\newcommand{\wb}{\omega_{\mathrm{b}}}
\newcommand{\wo}{\omega_0}
\newcommand{\w}{\omega}

\begin{document}

\title{Breather trapping and breather transmission in a DNA model with
an interface.}
\author{A Alvarez}
\affiliation {Grupo de F\'{\i}sica No Lineal. \'{A}rea de
F\'{\i}sica Te\'{\o}rica.
 Facultad de F\'{\i}sica. Universidad de Sevilla.
    Avda. Reina Mercedes, s/n. 41012-Sevilla (Spain)}
\author{JFR Archilla}
\author{J Cuevas}
\affiliation{Grupo de F\'{\i}sica No Lineal. Departamento de
Fisica Aplicada I. ETSI Inform\'{a}tica. Universidad de Sevilla.
Avda. Reina Mercedes, s/n. 41012-Sevilla (Spain)}
\author{PV Larsen}
\affiliation{Department of Mathematics, Technical University of
Denmark. DK-2800 Kgs. Lyngby (Denmark)}
\author{FR Romero}\email{romero@us.es}
\affiliation {Grupo de F\'{\i}sica No Lineal. \'{A}rea de
F\'{\i}sica Te\'{\o}rica.
 Facultad de F\'{\i}sica. Universidad de Sevilla.
    Avda. Reina Mercedes, s/n. 41012-Sevilla (Spain)}

\date{March 27, 2006}

\keywords {Moving breathers, breathers collisions, Klein--Gordon
lattices, dipole--dipole interaction, DNA dynamics}

\pacs { 63.20.Pw,
 63.20.Ry,
 63.50.+x,
  66.90.+r,
  87.10.+e}

\begin{abstract}

   We study the dynamics of moving discrete breathers in an interfaced piecewise DNA molecule.
 This is a DNA chain in which all the base pairs are identical and
 there exists an interface such that the base pairs dipole moments at each side are oriented in opposite
directions.
 The Hamiltonian of the Peyrard--Bishop model is augmented with a
term that includes the dipole--dipole coupling between base pairs.
Numerical simulations show the existence of two dynamical regimes.
If the translational kinetic energy of a moving breather launched
towards the interface is below a critical value, it is trapped in
a region around the interface collecting vibrational energy. For
an energy larger than the critical value, the breather is
transmitted and continues travelling along the double strand with
lower velocity. Reflection phenomena never occur.

 The same study has been carried out when a single dipole is oriented
 in opposite direction to the other ones.
  When moving breathers collide
with the single inverted dipole, the same effects appear. These
results emphasize the importance of this simple type of local
inhomogeneity as it creates a mechanism for the trapping of
energy.

 Finally, the simulations show that, under favorable conditions,
 several launched moving breathers can be trapped successively at the interface
 region producing an accumulation of vibrational energy. Moreover,
 an additional colliding moving breather can produce a saturation
 of energy and a moving breather with all the accumulated energy
 is transmitted to the chain.

\end{abstract}

\maketitle

\section{Introduction}

Nonlinear physics of DNA has experienced  an enormous development
in the previous years. There are many experimental data and
theoretical results published about the nonlinear properties of
DNA  (for a review see, e.g., ~\cite{YAK}.
 The DNA molecule is a discrete system consisting of many atoms
having a quasi-one-dimensional structure. It can be considered as
a complex dynamical system, and, in order to investigate some
aspects of the dynamics and the thermodynamics of DNA, several
mathematical models have been proposed. Among them, it is worth
remarking the Peyrard--Bishop model~\cite{PB89} introduced for the
study of DNA thermal denaturation. This model, and some variations
of it, has also been used extensively for the study of the
dynamical properties of DNA.

In DNA there exist different kinds of interactions between the
main atomic groups. One of them is the stacking interaction
between neighbouring bases along the DNA axis, these are
short--range forces which stabilize the DNA structure and hold one
base over the next one forming a stack of bases. There exist also
long--range forces, due to the finite dipole moments of the
hydrogen bonds within the nucleotides.
 The existence of these forces is corroborated both by theoretical
 and experimental studies. Recently, quantum chemical
 calculations, using the second Moler-Plesset perturbation method,
 have determined bonds lengths and bonds angles involved in
 hydrogen bonds between the bases of DNA base pairs, as well as
 dipole moments based at the MP2/6-31G(d) and MP2/6-31G(d,p) basis sets
 ~\cite{DA05}. The dipole moments calculated for the adenine-thymine base pair
 at equilibrium are, respectively, 1.44 D and 1.29 D. The values
 calculated for the guanine-cytosine base pair at equilibrium are,
 respectively, 5.88 D and 5.79 D. All these values are within the
 range of the parameters values for the dipole moments considered
 in this paper.

    Some experiments carried out on the dynamics of the B-A transition of
 DNA double helices, through the analysis of the transients
 observed under electric field pulses, show that dipolar
 strectching is the main driving force for the B-A reaction ~\cite{DD04}.

 The dipole--dipole interactions between different base pairs should never be ignored and they may
play a crucial role for the dynamical properties of DNA, for
example when the geometry of the double strand of DNA is taken
into account, as the distance between base pairs and, therefore,
the intensity of the coupling between them, depends on the shape
of the
molecule~\cite{GMCR97,MCGJR99,ACMG01,ACG01,CPAR02,LCBAG04,LCBAG004}.

 Most of the studies have been done considering DNA homogeneous models.
 Nevertheless, the DNA molecule is essentially an inhomogeneous system,
 and this inhomogeneity is characterized by the existence of different local
ranges, or functional regions, with very specific sequences of
base pairs and very specific functions.

Numerical simulations show that discrete breathers (DBs), which
are spatially localized nonlinear oscillations, can appear in
models of crystals, biomolecules, and many others nonlinear
discrete systems ~\cite{A97,FW98,M00}. It is well known that DBs
can be static, but they can also move under certain physical
conditions ~\cite{CAT96,AC98,Cretegny}, constituting a mechanism
for the transport of energy and information along discrete
systems. They are called \emph{moving breathers} (MBs) and have
been obtained in many different systems ~\cite{MER98,MER00,MRE01}.

 In the  Peyrard--Bishop model, the existence of DBs has
been demonstrated~\cite{DPW92,DPB93}, and DBs are thought to be
the precursors of the bubbles that appear prior to the
transcription processes in which large fluctuations of energy have
been experimentally observed~\cite{GKL87}.

 The study of MBs in Klein-Gordon models of
 DNA chains with the inclusion of long--range
interactions was initiated in~\cite{CAGR02} using an augmented
Peyrard--Bishop model. The system considered is homogeneous and
without bending, that is, all the base pairs are identical,
aligned in a straight line, and all the dipole moments are
parallel along the same direction. The results show that MBs exist
for a wide range of the parameter values, although the mobility is
hindered when the intensity of the long--range interaction
increases. Essentially, the effective mass of a breather increases
as the intensity of the dipole--dipole interaction increases, and
there exists a threshold of the dipole--dipole coupling constant
above which the breather is not movable.

The interaction of MBs with different kind of inhomogeneities in
the DNA molecule may be also of great importance for the
localization of energy~\cite{FPM94,MAS97}. As a consequence of
this interaction, MBs may be trapped in small regions, that is,
vibrations remain affecting only to a small number of consecutive
base pairs, producing an accumulation of energy, which can
initiate denaturation or transcription bubbles. Also, trapping
phenomena may be on the basis for the appearance of secondary
breaks when DNA is irradiated with ionizing radiation~\cite{B85}.

Structural studies carried out with non coding sequences of DNA
have shown that poly(dA.dT) sequences (called T-tracts) are
abundant genomic DNA elements. Suter et al.~\cite{SST00}have shown
that they exist as rigid DNA structures in nucleosome-free yeast
promoters in vivo, that is, in vivo they are not folded in
nucleosomes. Their data support that transcription activation
depends on the length of the T-tracts and the same occurs when
they are replaced by poly(dG.dC) sequences, another rigid
structures. This fact suggests that the T-tracts operate not by
recruitment of specific transcription factors, but rather by their
intrinsic DNA structure, and that transcription activation must be
intimately related with the dynamical properties of T-tracts.

These considerations motivated us to study MBs in some types of
poly(dA.dT) or poly(dG.dC) sequences. In this paper we study the
effects of the collisions of MBs with local inhomogeneities in two
different but related systems.

\begin{figure}
 \includegraphics[width=0.7\textwidth]{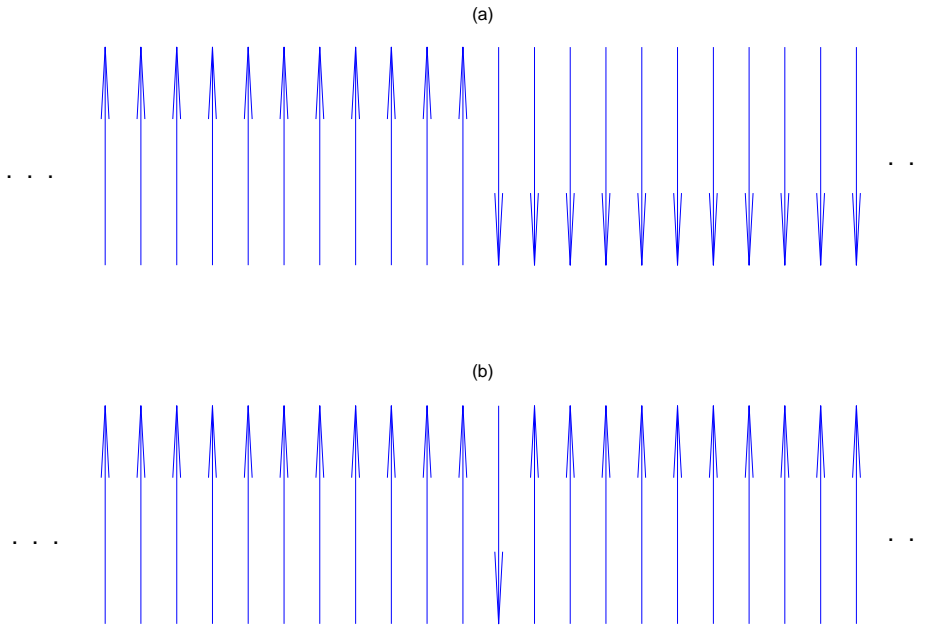}
 \caption{Sketch of the dipole chains.(a): the interfaced piecewise
  chain, (b): the single inverted--dipole chain}
 \label{Fig1}
\end{figure}

The first system consists of an inhomogeneous DNA molecule formed
by two consecutive homogeneous regions. All the base pairs are
identical, but the dipoles of a region are oriented in opposite
direction to the dipoles of the other one, as sketched in
Fig.~\ref{Fig1}(a). One possibility is for a sequence being as
$...AT/AT/AT/TA/TA/TA...$, where $AT$ and $TA$ represent,
respectively, the adenine-thymine base pair and the
thymine-adenine base pair. The other possibility is for a sequence
being as $...CG/CG/CG/GC/GC/GC...$, where $CG$ and $GC$ represent,
respectively, the cytosine-guanine base pair and the
guanine-cytosine base pair. The inhomogeneity is determined by the
existence of an interface that separates these regions, that is,
there exists a local inhomogeneity and the interface marks a
discontinuity in the system. These kind of chains are called
\emph{ interfaced piecewise chains}, they exist in real DNA and
can also be synthesized in the laboratory ~\cite{KOKMK71, MZK06}.

 The second system consists in a DNA chain of identical base pairs with all the
dipoles oriented in the same direction, except one of them which
has an opposite orientation, that is, the local inhomogeneity is
at the site of the inverted--dipole, as sketched in
Fig.~\ref{Fig1}(b).

In both systems, the simulations of MBs launched towards the
inhomogeneity show that if the translational kinetic energy is
below a critical value, which is determined by the values of the
coupling parameters, the breather is trapped. For greater values,
the breather is transmitted to the chain, losing part of its
energy as phonon radiation. Moreover, reflection of the colliding
breather never occurs.

These results emphasizes the enormous importance of a change of
the orientation of a single base pair in a homogeneous track of
DNA chain, as it demonstrates that a very simple structural change
creates a mechanisms for localization and accumulation of energy.

Moving breathers can also collide with a trapped breather. Our
study of these collision processes shows that multiple scenarios
are possible, one of them is the possibility of getting an
accumulation of energy at the interface up to a saturation
threshold value. All the accumulated energy can be carried away by
a successful incoming breather. These phenomena suggest a way for
the DNA molecule to accumulate enough energy to break the hydrogen
bonds between the base pairs on opposing strands.

Another important result concerning these systems, with a local
inhomogeneity and with competing short and long--range
interactions, is that there exists a relevant parameter
determining the range of translational kinetic values for which
MBs are trapped, this is the quotient between the dipole--dipole
and the stacking coupling constants. The critical kinetic energy
increases, or the size of the "trapping window", if this quotient
increases.

This paper is organized as follows: In section \ref{model}, we
introduce the interfaced piecewise DNA model and the single
inverted dipole DNA model, which includes the long--range
interactions and has into account their respective chain
conformations. In section \ref{linmod}, we investigate the
characteristics of the linear modes of the first system, as they
are important for determining possible behaviors of MBs when they
collide with the interface. It is shown that the linear modes
spectrum of this system is similar to the spectrum of a
homogeneous DNA molecule with a single inverted dipole. In section
\ref{numres}, we generate numerically MBs which are launched
towards the interface and we study the collision phenomena,
exploring all the possible scenarios that appear when the
translational kinetic energies of the MBs and the coupling
parameters are varied. Thereafter, we show that the collisions of
MBs in the case of a single inverted dipole DNA molecule, bring
about similar effects to the interfaced case. Section
\ref{collisions} presents some interesting phenomena that can
appear when moving breathers collide with a previous trapped
breather. Finally, section \ref{conclu} summarizes our results and
contains some conclusions.

\section{Interfaced piecewise DNA model}\label{model}

We consider a modification of the Peyrard-Bishop DNA model
~\cite{PB89}, with the addition of a dipole-dipole energy term.
Thus, the Hamiltonian of the system can be written
as~\cite{CAGR02}

\begin{equation}
\label{ham}
    H=\sum_{n=1}^N\left(\frac{1}{2}m\dot u_n^2+D(e^{-bu_n}-1)^2+
    \frac{1}{2}\varepsilon(u_{n+1}-u_n)^2+\frac{1}{2}\mu\sum_{m\ne n}J_{n,m}u_{m} u_n\right)
\end{equation}

The term $\frac{1}{2}m\dot u_n^2$ represents the kinetic energy of
the nucleotide of mass $m$ at the $n$th site of the chain, and
$u_n$ is the variable representing the transverse stretching of
the hydrogen bond connecting the base at the $n$th site. The Morse
potential, i.e., $D(e^{-bu_n}-1)^2$, represents the interaction
energy due to the hydrogen bonds within the base pairs, being $D$
the well depth, which corresponds to the dissociation energy of a
base pair, and $b^{-1}$ is related to the width of the well. The
stacking energy is $\frac{1}{2}\varepsilon(u_{n+1}-u_n)^2$, where
$\varepsilon$ is the stacking coupling constant. The last term of
the Hamiltonian, i.e., $\frac{1}{2}\mu\sum_{n}\sum_{m\ne
n}J_{n,m}u_{m} u_n,$ is the long-range dipole--dipole interaction
term, where $\mu$ is the dipole--dipole coupling constant. The
expression for this constant is $\mu=q^2/4\pi\varepsilon_0d^3$
~\cite{CAGR02}, $q$ being the charge transfer within the dipole
and $d$ the distance between neighbouring base pairs, which is
supposed to be constant. Finally, $J_{n,m}$ is the dipole-dipole
coupling factor given by
\begin{equation}\label{JJ}
    J_{n,m}=\frac{\alpha_{n,m}}{|m-n|^3},
\end{equation}
where $|m-n|$ is the normalized distance between base pairs, and
$\alpha_{n,m}$ takes the value $1$ if the dipoles at $n$ and $m$
are parallel, and $-1$ if they are antiparallel.

  In order to keep the spatial homogeneity in a finite system with
periodic boundary conditions~\cite{F98}, the number of particles
affected by the long--range interaction must be limited. In
consequence, if N is the number of base pairs, we suppose that the
long--range interaction affects to $(N-1)/2$ neighboring base
pairs, if N is odd, or $(N-2)/2$, if N is even, to each direction
of a given site of the chain. In fact, we use periodic boundary
conditions in a way that both the last dipole and the first one
are pointing along the same direction, this implies that the
interfaced piecewise chain must has the geometry of a M\"{o}bius
band, and in this way there exists only one interface in the
chain. As an example, we can consider a small system with $N=8$,
with the dipoles located at $n=+1,+2,+3,+4$ pointing in one
direction and the other ones located at $n=-1,-2,-3,0$ pointing in
opposite direction. Then, the dipole--dipole coupling factors
$J_{n,n+p}$ can be written in a matrix form as follows:\\

\begin{equation}
\left( \begin{array}{cccccccc}
  0_{}      & 1_{}      & 0.1250    & 0.0370        & 0_{}       & 0.0370_{} & 0.1250_{} & 1_{} \\
  1_{}      & 0_{}      & 1_{}      & 0.1250_{}     & -0.0370_{} & 0_{}      & 0.0370_{} & 0.1250_{} \\
  0.1250_{} & 1_{}      & 0_{}      & 1_{}          & -0.1250_{} & -0.0370_{}& 0_{}      & 0.0370_{} \\
  0.0370_{} & 0.1250_{} & 1_{}      & 0_{}          & -1_{}      & -0.1250_{}& -0.0370_{}& 0_{} \\
  0_{}      & -0.0370_{}& -0.1250_{}& -1_{}         & 0_{}       & 1_{}      & 0.1250_{} & 0.0370_{} \\
  0.0370_{} & 0_{}      & -0.0370_{}& -0.1250_{}    & 1_{}       & 0_{}      & 1_{}      & 0.1250_{} \\
  0.1250_{} & 0.0370_{} & 0_{}      & -0.0370_{}    & 0.1250_{}  & 1_{}      & 0_{}      & 1_{} \\
  1_{}      & 0.1250_{} & 0.0370_{} & 0_{}          & 0.0370_{}  & 0.1250_{} & 1_{}      & 0_{}%

\end{array}
 \right)\nonumber
\end{equation}

We perform the following changes of variables:
\begin{equation}\label{change}
    t\rightarrow \wo t, \,  u_n\rightarrow bu_n, \,
    \varepsilon\rightarrow \frac{\varepsilon}{m\wo^2},\,
    \mu\rightarrow \frac{\mu}{m\wo^2}, \,  H\rightarrow \frac{H}{2D},
    \,  D\rightarrow \frac{1}{2},
\end{equation}

where  $\wo=\sqrt{2b^2D/m}$ is the frequency of an isolated
oscillator in the harmonic limit. It becomes unity in the scaled
system.

With these changes, the dynamical equations become:

\begin{equation}\label{F}
\ddot u_n+(e^{-u_n}-e^{-2u_n})+\varepsilon(2u_n-u_{n+1}-u_{n-1})+
\mu\sum_{m\ne n}\frac{\alpha_{n,m}}{|m-n|^3}u_{m}=0.
\end{equation}

\section{Analysis of the linear modes}\label{linmod}

The study of the linear modes (phonons) of a nonlinear discrete
system gives a necessary information for predicting some
properties of the MBs with a given frequency. For example, the
frequencies of MBs must be not too close to the phonon band.
Otherwise, they emit a large amount of phonon radiation
simultaneously to their movement. Furthermore, as it was shown in
\cite{CPAR02b}, the analysis of linear modes allows to predict the
scenario of the collisions of MBs with inhomogeneities without the
necessity of performing numerical simulations.

The dynamical equations of the interfaced piecewise system can be
linearized if we assume that the amplitudes of the oscillations
are small enough. Then, the linearized dynamical equations are:

\begin{equation}\label{Flinear}
    \ddot u_n+u_n+\varepsilon(2u_n-u_{n+1}-u_{n-1})+
    \mu\sum_{p\ne 0}J_{n,n+p}u_{n+p}=0.
\end{equation}

The inhomogeneity of these equations determines the existence of
two localized modes, one of them is above (top mode) and the other
one is below (bottom mode) the extended modes (phonon band).

\begin{figure}
\includegraphics[width=0.7\textwidth]{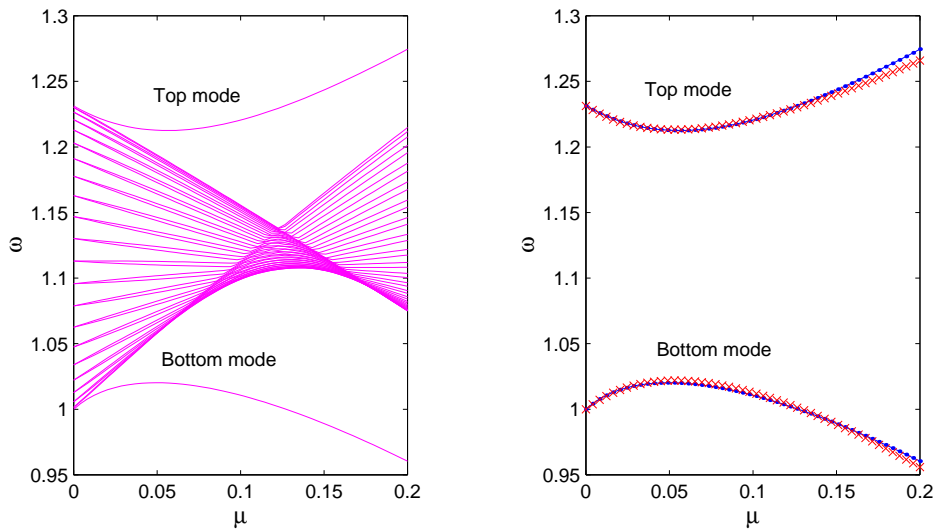}
 \caption{Left: Frequencies of the linear modes varying the coupling parameter $\mu$ for
    $\varepsilon=0.129$. The upper line represents the top mode
    and the lower line represents the bottom mode. Right: Top and bottom mode frequencies with
   respect to
     $\mu$. Crosses and dots represent, respectively, analytical and numerical values.}
    \label{Fig2}
\end{figure}

We have calculated numerically the frequencies of the linear modes
for different values of the parameters $\varepsilon$ and $\mu$.
Fig.~\ref{Fig2}(left) shows the dependence of the frequency
spectrum with respect to the parameter $\mu$ for the fixed value
of $\varepsilon=0.129$, which is an appropriate value to obtain
MBs with low dispersion~\cite{CPAR02}. In this figure, it can be
observed the existence of localized modes whose origin relies in
the fact that the interface acts as an inhomogeneity in the
lattice. The localized modes are two, one of them is above (top
mode) and the other one is below (bottom mode) the extended modes
(phonon band).

\

\begin{figure}
        \includegraphics[width=0.7\textwidth]{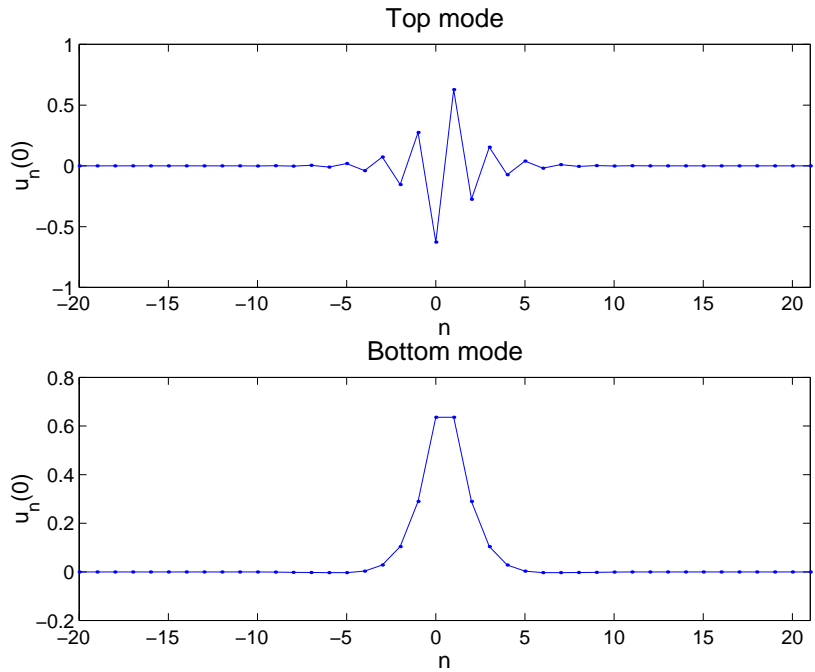}
    \caption{Generic profiles of the top and bottom localized modes for the coupling
values $\mu=0.05$ and
    $\varepsilon=0.129$.}
    \label{Fig3}
\end{figure}

 The generic profiles of the localized modes are shown in
Fig.~\ref{Fig3}. Their vibration patterns depend on the parameter
$\mu$. In particular, we have observed that there exists a
critical value $\mu_c$ such that for $\mu<\mu_c$, the top mode has
a zig-zag vibration pattern, while the bottom mode vibrates in
phase. For $\mu>\mu_c$, the vibration patterns are interchanged.
As it is shown below, $\mu_c\approx\varepsilon$ (e.g. for
$\varepsilon=0.129$, $\mu_c=0.123$). This fact is related to the
existence of a competition between the attractive (stacking) and
the repulsive (dipole--dipole) interaction. The first one is
dominant for $\varepsilon\gtrsim\mu_c$ and vice versa.

An approximate explicit expression that gives the dependence of
the localized modes frequencies with respect to the coupling
constant $\mu$ can be obtained considering that the long-range
interaction is limited to nearest neighbour base pairs. A study of
infinity-range interaction could be performed, but it would be
limited to frequencies close to the phonon band~\cite{F98,GMCR97}.
This approximation is justified as the interaction decays rapidly.
We have checked that the results are practically  coincident
except for values of $\mu$ far from the ones considered in our
study.

The localized modes can be found using Green's lattice function
methods~\cite{BK91,DPW92,MMW}. However, the same results can be
obtained in a more straightforward way using the following ansatz:

\begin{equation}
\left\{
    \begin{array}{lll}
    u_n=a_0r^{|n|}\exp(\mathrm{i}\,\omega \,t)  & \textrm{for} & n>0 \\
    u_n=a_1r^{|n-1|}\exp(\mathrm{i}\,\omega \,t) &  \textrm{for}& n<1 \\
    \end{array},
\right.
\end{equation}
where r is a spatial decay factor. The sign of r indicates the
vibration pattern of the localized mode. If $r>0$, the particles
vibrate in phase, and if $r<0$, the mode has a zig-zag vibration
pattern.

We have supposed that the dipoles at $n\ge 1$ are all antiparallel
to the dipoles at $n\le 0$, then, there exists only two
neighboring dipoles which have a different orientation, and the
interface is located between $n=0$ and $n=1$. As the localized
modes are centered between both (see Fig.~\ref{Fig3}), we assume
that:

\begin{equation}
\left\{
    \begin{array}{ll}
    a_0=-a_1 &\textrm{(corresponds to the top mode)} \\
    a_0=a_1 & \textrm{(corresponds to the bottom mode)} \\
    \end{array}.
\right.
\end{equation}

With the application of these ans\"{a}tze we obtain that the
frequencies of the top and bottom modes are, respectively:

\begin{equation}
    \w_{\mathrm{top}}^2=\wo^2+2\frac{\mu^2+2\varepsilon^2+\varepsilon\mu}
    {\varepsilon+\mu}
\quad;\quad
\w_{\mathrm{bottom}}^2=\wo^2+2\mu\frac{\varepsilon-\mu}{\varepsilon+\mu}.
\end{equation}

The decay factors for the top and bottom modes are, respectively:

\begin{equation}
    r_{\mathrm{top}}=\frac{\mu-\varepsilon}{\mu+\varepsilon}\quad;\quad
    r_{\mathrm{bottom}}=-r_{\mathrm{top}}.
\label{eq:rtopbottom}\end{equation}

>From these expressions, it is easy to obtain that
$\mu_c=\varepsilon$. The discrepancies from this value and the
numerical one rely on our assumption that long-range interactions
are limited to nearest neighbours.
 The values of the
parameters for which the breather frequency
$\wb=\w_{\mathrm{bottom}}$ are related to the scattering
properties, see Ref.~\cite{CPAR02b}. However, these values are
excluded in our study, because as we require the existence of
moving breathers, we need that $\mu<M$ and $\varepsilon>M$, with
$M\approx0.2\wb^{2.4}$~\cite{CAGR02}. In other words,
$\varepsilon>\mu$ is a necessary condition for the existence of
moving breathers. The equality $\wb=\w_{\mathrm{bottom}}$ implies
that $\mu=[(2\varepsilon-\delta)+\sqrt{4\epsilon^2+\delta^2
-8\varepsilon\delta}]/2$, with $\delta\equiv\wb^2-\wo^2<0$. Thus,
it is straightforward to demonstrate that $\mu$ must be greater
than $\epsilon$ in order to get breather resonance with the bottom
mode frequency. This region is inaccessible for the moving
breathers, as it is shown above.

The analytical and numerical results that give the dependence of
the frequencies of the localized modes with respect to $\mu$ are
represented in Fig.~\ref{Fig2}(right). As can be appreciated,
there is an excellent agreement between the two approaches.

The single inverted dipole system also has two linear localized
modes, but now they are localized at the inverted dipole site. The
frequencies of these modes are:

\begin{equation}
    \w_{\mathrm{inv}}^2=\wo^2+2\varepsilon\pm\frac{2\varepsilon^2-4\varepsilon\mu-14\mu^2}
    {\sqrt{\varepsilon^2+6\varepsilon\mu-15\mu^2}}.
\end{equation}

\section{Evolution of moving discrete breathers in the interfaced piecewise
DNA chain.}\label{numres}

  The dynamics of MBs can be strongly affected by the existence of
the interface. It is expected that MBs differing only in their
translational kinetic energies can produce different effects when
they interact with the interface. In order to study these effects,
we have simulated many collisions processes launching MBs towards
the interface with different translational kinetic energies.

  MBs can be generated numerically by adding to the velocities of a static breather a
perturbation of magnitude $\lambda$ collinear to the pinning
mode~\cite{CAT96,AC98}. The pinning mode is an anti--symmetric
linear localized mode, which may appear in the set of linear
perturbations of the system when the coupling is strong enough.
This perturbation breaks down the translational symmetry of the
system and the breather moves with a translational kinetic energy
given by $K=\lambda^2/2$. The time evolution of the breathers have
been studied using a Calvo's 5th order symplectic
algorithm~\cite{SC94}.

 Recent quantum--mechanical calculations Ref.~\cite{CSAH04} establishes
that in DNA the order of magnitude of the parameter $\mu$ is
$0.002$, therefore, as we are interested in applications to DNA,
we have limited our study to this order.

  MBs are not exact solutions of the dynamical equations and some phonon
radiation is emitted simultaneously to the movement of breathers.
This radiation is minimized if the breather frequency is not close
to the phonon band. In all the simulations, the breather frequency
has been fixed to the value of $\wb=0.8$, which is an appropriate
value. For smaller values of $\wb$, the breather gets more and
more sharply localized and highly pinned to the lattice, and for
$\wb\lesssim 0.67$ its movement is not possible. Also, phonon
radiation increases when the value of the translational kinetic
energy increases and some effects due to the size of the system
can appear. For that reason the calculations have been tested
using different systems sizes, to make sure that the results are
not modified by boundary effects (most of the simulations have
been done with $N>200$). The values of the parameter $\lambda$
have been limited to the interval $(0.01,0.3)$, as for this
interval the phonon radiation is small enough. Finally, the first
group of simulations have been done with the value of the stacking
coupling constant $\varepsilon=0.129$.

We have performed a large number of simulations and, essentially,
we have found that there exist only two different dynamical
regimes, which can be characterized as follows:

\begin{description}

\item[Trapping regime]:

If the translational kinetic energy of a MB is smaller than a
critical value, i.e. $\lambda<\lambda_c$, the breather is trapped
at the discontinuity region, that is, only some particles around
the interface remain oscillating. The critical translational
kinetic energy depends on the values of the parameters $\mu,
\epsilon$ and $\omega_b$. An example of the time evolution of a MB
with $\lambda$ smaller than the critical value, is observed in
Fig.~\ref{Fig4}(a). However, not all the moving breather energy is
trapped after the collision because some energy is transferred to
the chain as phonon radiation. To see this, we have studied the
evolution of the "central energy", i.e. the energy of some few
particles around the interface, which is represented in
Fig.~\ref{Fig5}(a). Before the collision the central energy is
zero, then it increases quickly taking the value of the incident
breather energy, and a small decay of this energy occurs which is
due to phonon radiation. This can be appreciated in the central
zoom figure of Fig.~\ref{Fig5}(a).
\begin{figure}
        \includegraphics[width=0.7\textwidth]{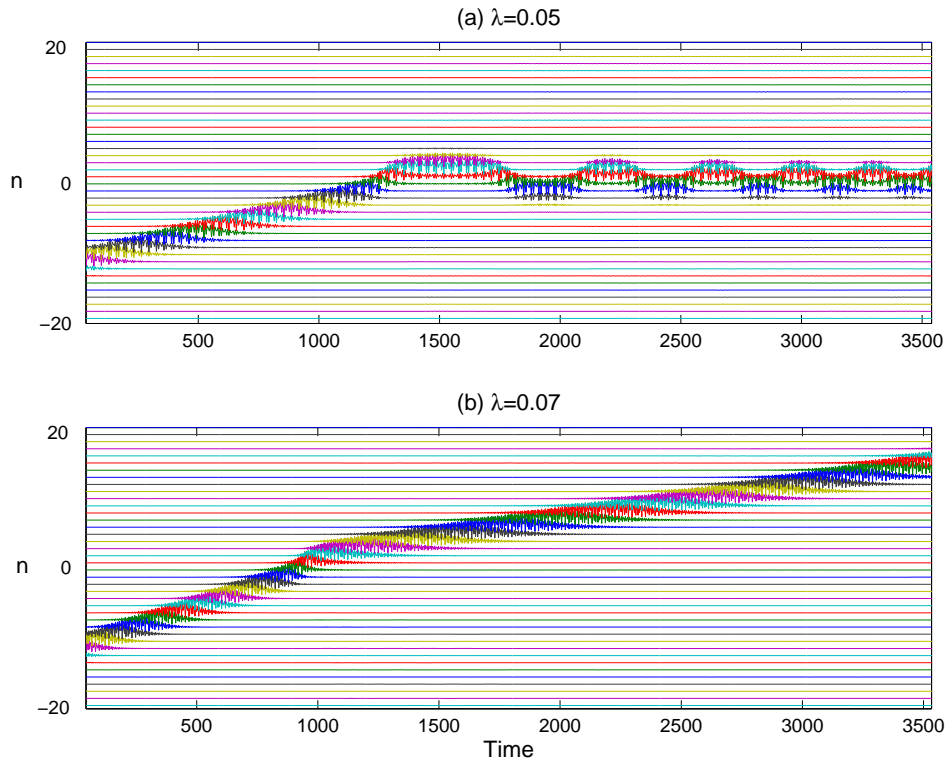}
    \caption{ Two different regimes for the interaction of MBs with the
    interface.
    Displacement from the n-th
    equilibrium position versus time, for two different values of the translational kinetic
    energy: (a) $\lambda<\lambda_c$ and (b) $\lambda>\lambda_c$,
    with $\mu=0.002$, $\varepsilon=0.129$ and
    $\wb=0.8$. For these parameters values, $\lambda_c=0.056$. The interface is located between $n=0$ and $n=1$.}
    \label{Fig4}
\end{figure}

\begin{figure}
        \includegraphics[width=0.9\textwidth]{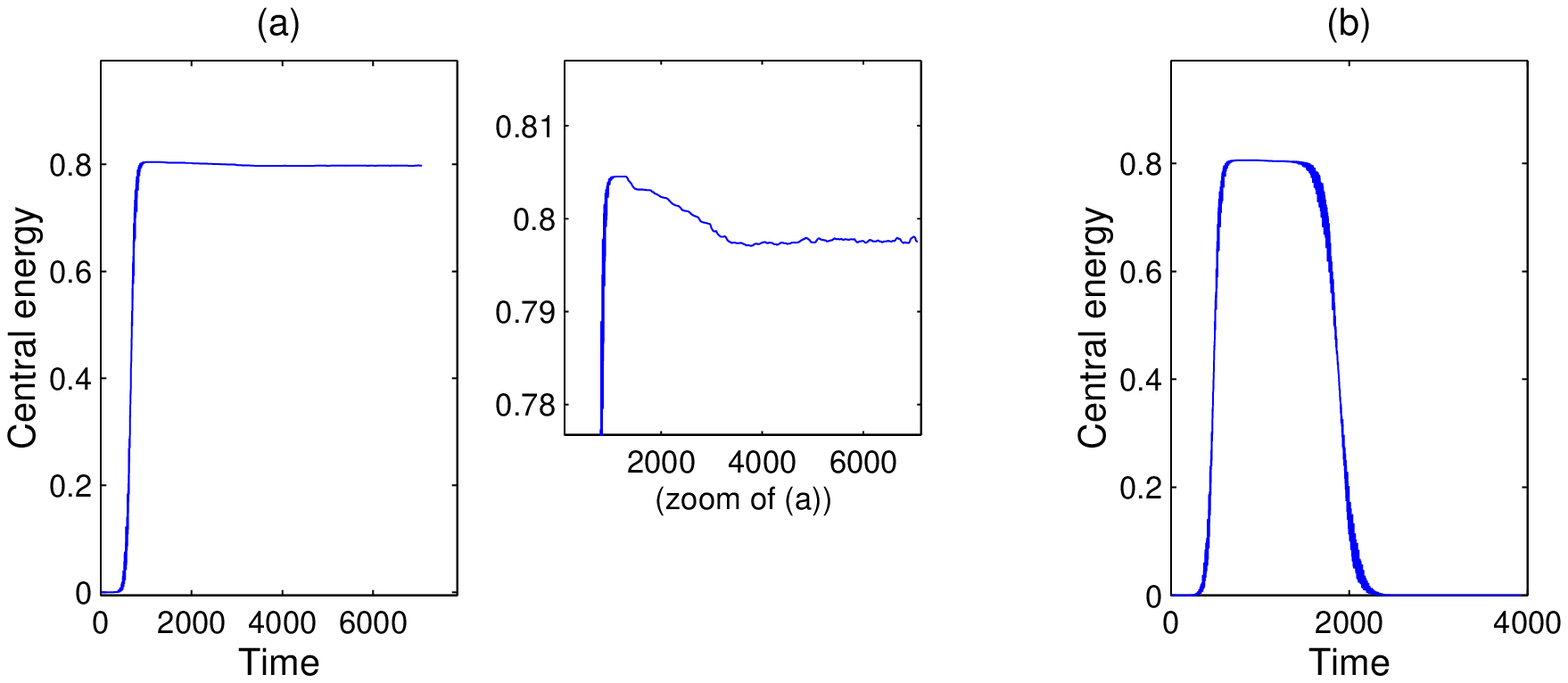}
    \caption{Time evolution of the central energy corresponding to
    the collision processes: (a) in Fig.~\ref{Fig4}(a); (b) in Fig.~\ref{Fig4}(b). The central figure is a zoom of
    the left one}
     \label{Fig5}
\end{figure}

The analysis of the Fourier spectrum of this trapped breather,
carried out after the initial decay of the central energy and at
an early stage of the evolution, shows a frequency of value 0.8,
which is the internal frequency $\omega_b$ of the launched
breather, and a frequency of value 0.02, which corresponds to an
oscillating movement of the breather around the interface. This
type of movement can be appreciated in Fig.~\ref{Fig4}(a). The
absolute values of the Fourier components calculated soon after
the collision of the breather with the interface are represented
in Fig.~\ref{Fig6}, and the central energy of this breather has
the value $0.7975$. The same calculations carried out after a time
of $200$  breather periods, shows a decrease of the Fourier
component corresponding to $\omega=0.02$, and a central energy of
value $0.7943$. A very slow--decaying process occurs and the
oscillating trapped breather approaches to a static breather
centered at the interface with frequency $\omega_b=0.8$. The
energy of this existing static breather is $0.7646$. Then, a
necessary condition for the trapping effect is the existence of an
\emph{inhomogeneity breather solution} (IB) with the frequency of
the MB and with a smaller energy. It consists of a static breather
solution centered at the inhomogeneity to which the trapped
breather decays.

\begin{figure}
        \includegraphics[width=0.6\textwidth]{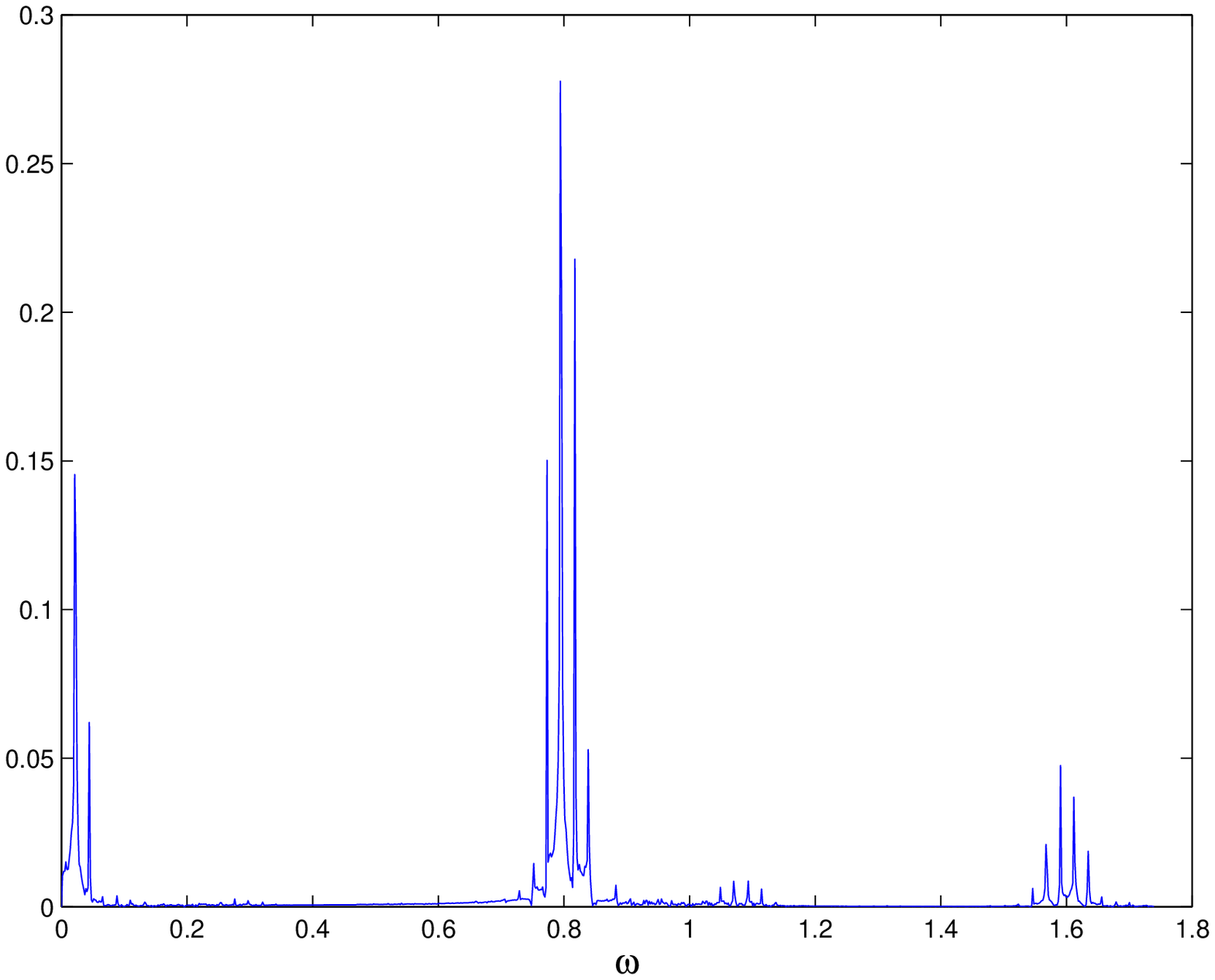}
    \caption{ Absolute values of the Fourier components of the trapped breather shown in Fig.~\ref{Fig4}(a)
    calculated soon after the collision}
     \label{Fig6}
\end{figure}

The trapping phenomena for small enough translational kinetic
energies, and the oscillatory behavior as a whole of the trapped
breather can be partially explained with the help of the
collective coordinate method.

In fact, the long--range interaction term of the Hamiltonian may
be represented as follows:
\begin{equation}\label{L}
 \frac{1}{2}\mu\sum_{n}\sum_{m\ne n}J_{n,m}u_{n}
   u_{m}= \sum_{n} V_{n}^{Eff}u_{n}^2-\frac{1}{4}\mu\sum_{n}\sum_{m\ne n}
   J_{n,m}(u_{n}-u_{m})^{2},
\end{equation}
 where the first summation term in the rhs represents an
effective on-site energy ~\cite{GMC00}, with the on-site potential

\begin{equation}\label{E}
 V_{n}^{Eff}=\frac{1}{2}\mu\sum_{m\neq n}J_{n,m}
\end{equation}
  and the second summation term represents an effective dispersion
  energy. In the case of a spatially homogeneous system the
  on-site potential does not depend on the index n, but for the
  interfaced piecewise system the potential $V_{n}^{Eff}$, with $\mu=0.002$, has the well
profile shown in Fig.~\ref{Fig7} (dots). It has a symmetric
profile, and the depth is close to $\mu$. The DNA chain with a
point inhomogeneity given by a single inverted--dipole
 has an effective potential with the well profile shown in Fig.~\ref{Fig7} (open circles).
 It is also a symmetric profile, and the depth is close to
$2\mu$.

The set of dynamical equations derived from the Hamiltonian
~(\ref{ham}) can be transformed, in the continuum limit, into the
corresponding continuous partial differential equation. Within
this context the collective coordinate method provides a good way
to analyze the influence of a perturbation on a soliton (in our
case we can consider the long range interaction term as a
perturbation). The idea is the same as using the center of mass to
analyze the behavior of a system of particles. The method
introduces the center of the breather "x" as a description
variable, and the on-site potential affects as a potential energy
of the form
\begin{equation}\label{VC}
 V(x)=\frac{1}{2}\mu\sum_{n}V_{n}^{Eff}u_{n-x}^2.
\end{equation}
If the breather is well localized, as occurs in our cases,
$V(x)\approx V_{x}^{Eff}$. In other words, the translational
movement of the breather center "x" is affected by a symmetric
potential well with the same form as the shown, for each case, in
Fig.~\ref{Fig7}. If the translational kinetic energy of the
breather is small enough, the phonon radiation emitted as it
passes through the potential well prevents the transmission of the
breather and it is trapped with an oscillating movement like a
particle in a potential well.
 The remaining terms in Eq.~(\ref{L}) also contribute to the
 center of breather motion but they are not so significant. In fact, in the
 continuum approximation they lead to a space-dependent effective
 mass of excitation.
\begin{figure}
 \includegraphics[width=0.4\textwidth]{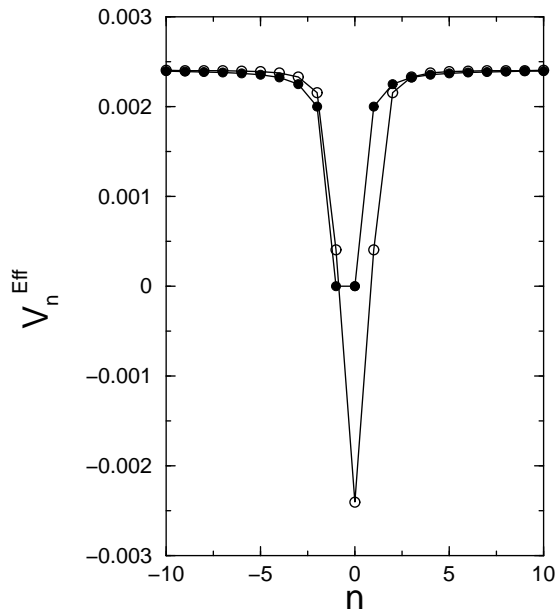}
 \caption{Effective potential $V_{n}^{Eff}$, versus site number,
 n, for the interfaced piecewise system (dots) and for the single inverted dipole system (open circles).}
 \label{Fig7}
\end{figure}

\item[Transmission regime]:

If the translational kinetic energy of a MB is bigger than the
critical value, i.e. $\lambda>\lambda_c$, the breather crosses the
interface and continues its movement along the chain with smaller
translational kinetic energy. In other words, the breather is
transmitted as a consequence of the interaction with the
interface, and a small part of its energy is transferred to the
chain as phonon radiation. An example of a transmitted MB can be
seen in Fig.~\ref{Fig4}(b), and the evolution of the corresponding
central energy is shown in Fig.~\ref{Fig5}(b).
\end{description}

The same qualitative results have been found in the case of the
parallel-oriented dipole chain with a single inverted--dipole. In
this case, the trapping site is centered at the inverted dipole.

Both this system and the interfaced piecewise system have a
symmetric effective on--site potential well due to the dipole
coupling. Also, both of them have two linear localized modes. The
nonexistence of an effective on--site potential barrier makes
impossible the appearance of a reflection regime.

\begin{figure}
         \includegraphics[width=0.4\textwidth]{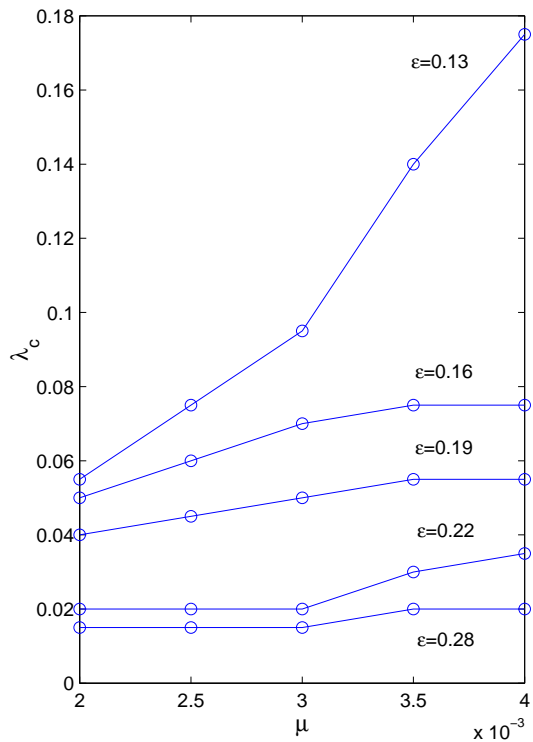}
    \caption{Dependence of $\lambda_c$ with respect to $\mu$ for some different values of
    $\varepsilon$.}
    \label{Fig8}
\end{figure}

As mentioned before, the critical value $\lambda_c$ depends on the
values of the parameters $\mu$, $\varepsilon$ and the breather
frequency $\w_b$. We have numerically determined the dependence of
$\lambda_c$ with respect to $\mu$ for some different values of
$\varepsilon$ and the fixed value $\omega_b=0.8$. The results are
shown in Fig.~\ref{Fig8}, where it can be observed that, for a
fixed value of $\varepsilon$, $\lambda_c$ increases monotonously
with $\mu$, and, for a fixed value of $\mu$, $\lambda_c$ decreases
monotonously with $\varepsilon$.

Equation ~(\ref{eq:rtopbottom}) shows that the extent of
localization of the linear localized modes depends on the values
of the parameters $\mu$ and $\varepsilon$. In fact, for the
interfaced system the decay factor for the bottom mode is given
approximately by

\begin{equation}\label{eq:rbottom}
  r_{\mathrm{bottom}}=(1-\frac{\mu}{\varepsilon})^{2}.
\end{equation}

Then, the extent of localization depends basically on the quotient
$\kappa={\mu}/{\varepsilon}$. We have observed that varying $\mu$
and $\varepsilon$ with constant $\kappa$, the values of
$\lambda_c$ are practically coincident. Also, when $\kappa$
increases, or $r_{\mathrm{bottom}}$ decreases, $\lambda_c$
increases. The IB solution is even more localized than the bottom
mode because $\w_b < \w_{\mathrm{bottom}}$. The values of the
critical translational kinetic energies that determine the
transitions from the trapping to the transmission regimes are
related to the extent of localization of the IB solution. An
important result is that for a given system the "trapping window",
defined as the range of translational kinetic energy for which
trapping occurs, is related to the quotient $\kappa$. Higher
localization hinders the capability of breathers for being
transmitted to the chain.

A similar conclusion can be obtained for the single inverted
dipole chain. In this case the values of $\lambda_c$ are higher
than the corresponding to the interfaced chain. For example, with
$\mu=0.002$ and $\epsilon=0.129$ we obtain $\lambda_c=0.195$. With
the same value of $\epsilon$ and with $\mu=0.0025$, $\lambda_c
>0.3$, although this value is out of the interval of $\lambda$ values that we have
considered in this paper.

The problem of the interaction of MBs with an impurity in a
homogeneous Klein-Gordon chain was considered in ~\cite{CPAR02b}.
In that case MBs can be reflected, trapped or transmitted by the
impurity, and a necessary condition for the appearance of trapping
is the existence of an impurity breather solution, that is, a
static breather solution centered at the impurity. Similarly, we
have seen that for our systems a necessary condition for trapping
is the existence of an \emph{inhomogeneity breather solution} (IB)
with the frequency of the MB and an internal energy smaller than
the internal energy of the MB.

\begin{figure}

         \includegraphics[width=0.7\textwidth]{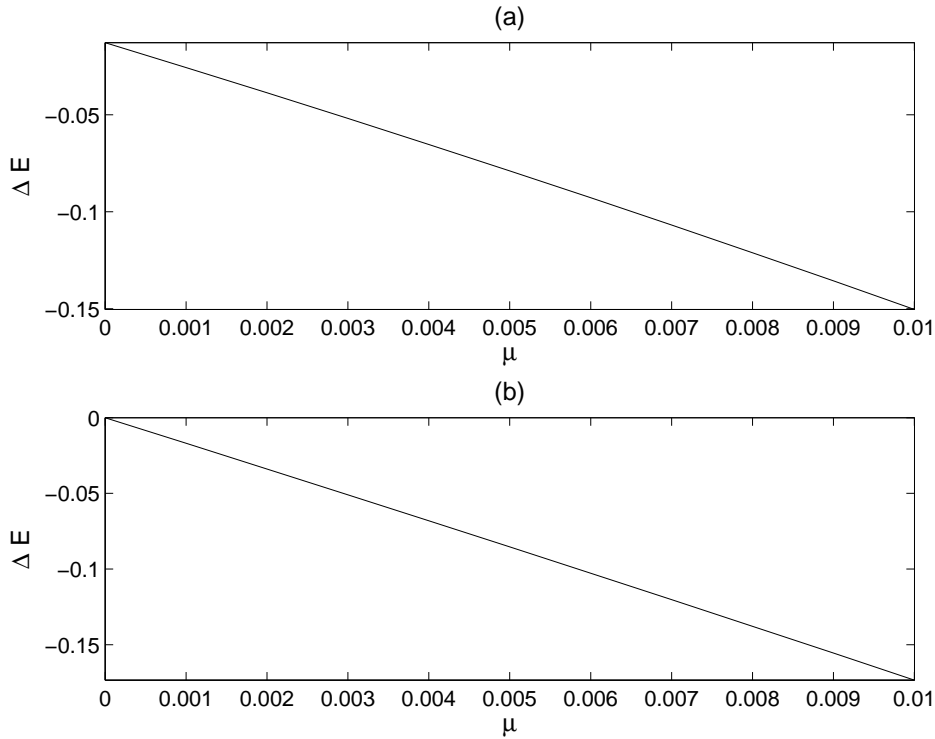}

    \caption{Dependence of the difference between the
    inhomogeneity
     breather energy ($E_{IB}$) and the internal energy of the moving breather ($E_{MB}$)
    with respect to $\mu$, for $\epsilon=0.129$ and $\omega_b=0.8$.
    (a) the interface case, (b) the single inverted--dipole case}
    \label{Fig9}
\end{figure}

Fig.~\ref{Fig9} shows that, for $\epsilon=0.129$ and
$\omega_b=0.8$, the difference between the inhomogeneity breather
energy ($E_{IB}$) and the internal energy of the moving breather
($E_{MB}$) is negative. This confirms the existence of
inhomogeneity breathers and that its energy is smaller than the
moving breather energy, which is the necessary condition for
trapping.

It can also be observed in this figure that the magnitude the
difference between the inhomogeneity breather energy ($E_{IB}$)
and the internal energy of the moving breather ($E_{MB}$)
increases monotonously with $\mu$. Thus, the phonon radiation
emitted by a trapped breather increases with $\mu$.

\section {Energy accumulation at the interface region: Breather collisions.}
\label{collisions}

The study of the trapping of several breathers at the interface
region is of great interest. When a MB is trapped, other incoming
breathers can also be trapped after colliding with the interface.
This process could increase the energy density locally, thus
collecting or accumulating energy in the interface region. Another
point of interest is whether there is a saturation threshold for
this accumulated energy .

 In general, the study of energy exchange in
collisions of MBs is a difficult problem. Some results are known
for a special type of FPU lattice ~\cite{Doi03}, and in a
one--dimensional DNA model ~\cite{TP96}, but at present no general
results are known concerning to collisions of MBs in Klein--Gordon
lattices.

 The simulations show that when a MB collides with a trapped breather, the effects of the
collision are strongly dependent on their exact dynamical details
when the effective collision processes begin. We have performed a
large number of simulations in a chain of N=200 base pairs, with
the coupling parameters $\mu=0.002$ and $\epsilon=0.129$. All the
breathers have the frequency $\omega_b=0.8$. We have found that
under favorable conditions the accumulation of energy is possible
as a consequence of the trapping of successive breathers.

\begin{figure}
  \begin{center}

    \includegraphics[width=0.75\textwidth]{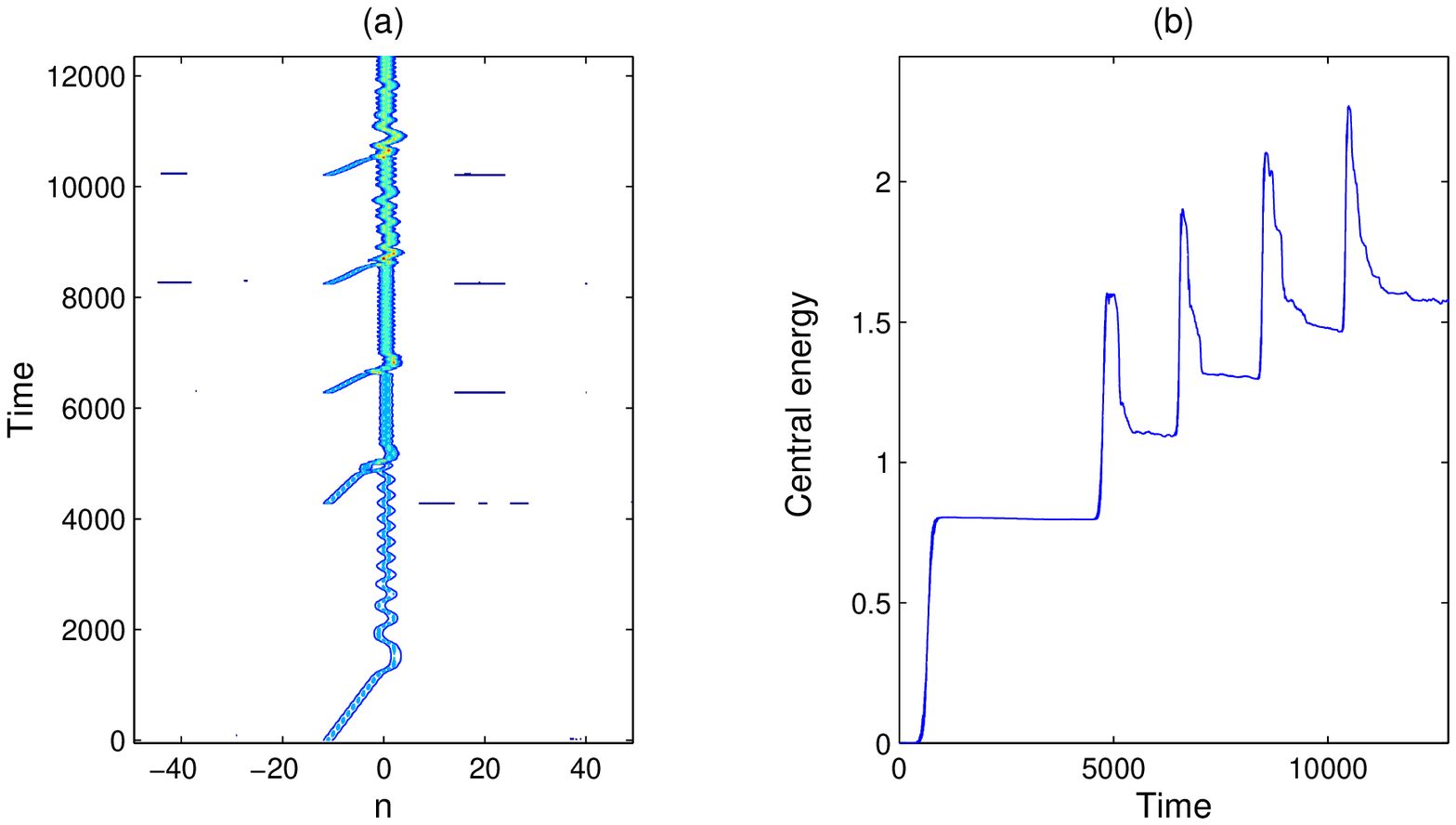}
   \end{center}
  \caption{
  (a) Contour plots for the evolution of five breathers launched towards the interface with different translational kinetic
  energies. The breathers are trapped at the interface increasing the
  accumulated energy density ($\lambda=0.05; 0.07; 0.15; 0.19; 0.2$).
  (b) Time evolution of the central energy.}
  \label{Fig10}
\end{figure}

We show in Fig.~\ref{Fig10}(a) contour plots for the evolution of
the energy density of five breathers launched towards the
interface at different instant of time. All the translational
kinetic energies are different.

\begin{figure}
  \begin{center}

    \includegraphics[width=0.75\textwidth]{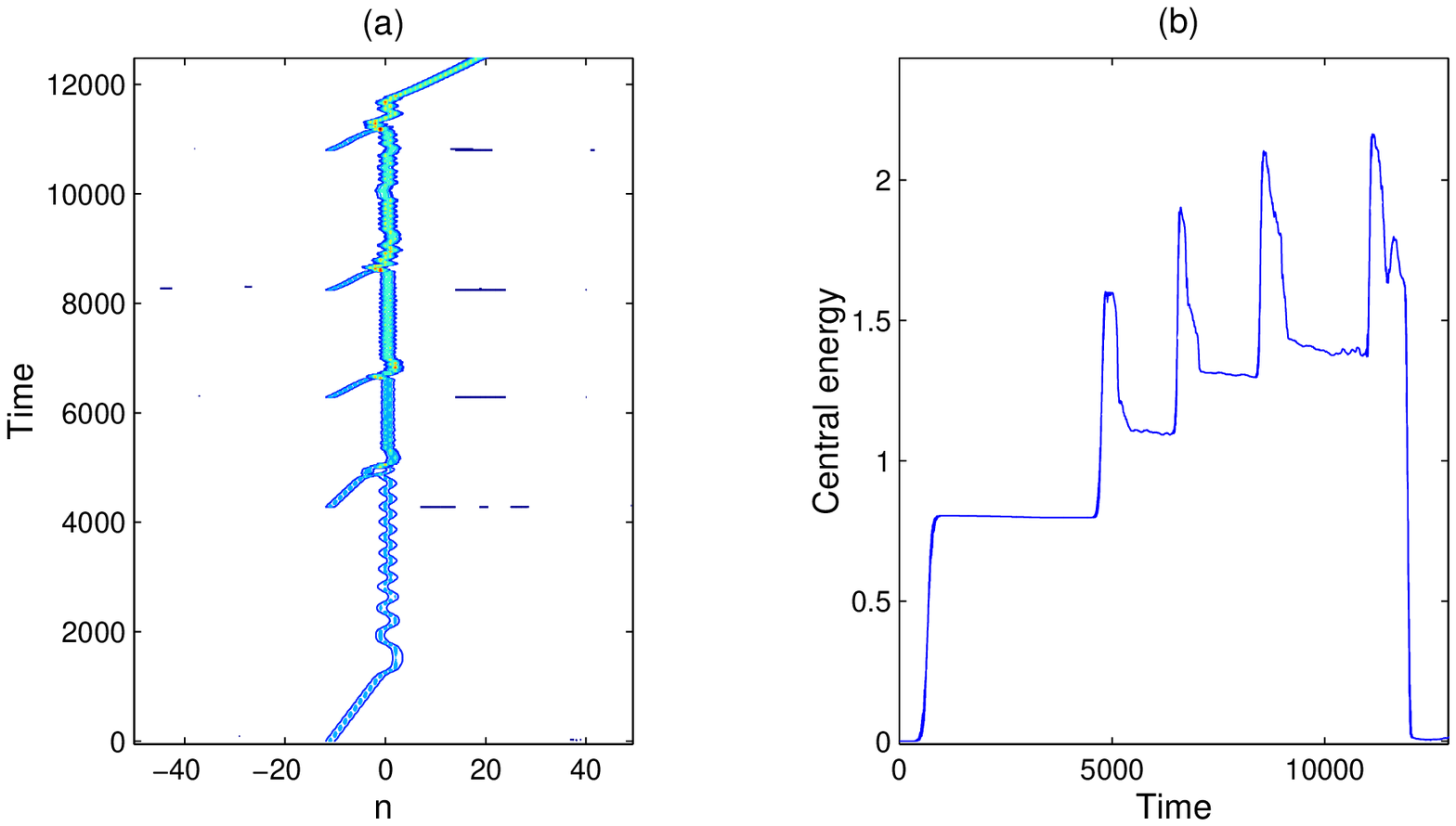}
   \end{center}
  \caption{
  (a) Contour plots for the evolution of five breathers launched towards
   the interface, with different translational kinetic energies
  ($\lambda=0.05; 0.07; 0.15; 0.18; 0.14$).
  After the last collision the accumulated energy is transmitted to the right side of the chain.
  (b) Time evolution of the central energy.}
  \label{Fig11}
\end{figure}

The first breather is launched towards the interface at $t=0$ with
a translational kinetic energy smaller than the critical value,
$(\lambda_c=0.056)$, so that the breather is trapped at the
discontinuity region.

 The phonon radiation must be removed because it distorts the
numerical simulations when it is reflected at the boundaries. For
that purpose, just before each breather is launched, we set to
zero the displacements and velocities outside the interface
region, $(|n|>5)$.

 The second breather is launched with $\lambda=0.07$, which is
greater than $\lambda_c$, but it is also trapped at the interface.
We have observed that the results of the collision between these
two breathers depends strongly on their relative phase when the
collision processes begin. We have chosen the translational
kinetic energy of this second breather in order to obtain an
efficient trapping.

Using the same procedure as before, three consecutive breathers
are launched with $\lambda=0.15; 0.19; 0.2$, respectively. All of
them are trapped, and to illustrate the evolution of the total
collected energy at the interface, we have represented in
Fig.~\ref{Fig10}(b) the central energy versus time. We observe
that after each collision the total trapped energy increases, but
only part of the energy of each incident breather is accumulated
at the interface region. The numerical simulations show that the
breather is partly transmitted and partly reflected, and also some
little  phonon radiation appears. It is interesting to observe
that the phonon radiation increases as the number of trapped
breathers or the energy accumulated increases. The possibility for
trapping more energy at the inhomogeneity region by additional
incoming breathers seems to have a limit due to a saturation
effect.

The accumulation of energy at the interface region could start
local base pair openings in the DNA molecule.
 Nevertheless, the simulations show that the accumulation of energy can,
occasionally, disappears after an appropriate collision. An
example of this is shown in Fig.~\ref{Fig11}, where the last
collision induces the transmission of the accumulated energy to
the right side of the chain. In this simulation five breathers are
launched successively towards the interface with $\lambda=0.05$,
$\lambda=0.07$, $\lambda=0.15$, $\lambda=0.18$ and $\lambda=0.14$,
respectively.

To gain deeper insight into the possible reasons for these
striking features, the authors are currently investigating this
issue and will be object of a future work.
\section{Conclusions}\label{conclu}

  We have studied the evolution, and some collision processes, of MBs in two different DNA molecules.
  The first DNA chain consists of a sequence of identical base pairs, where there
exists an interface such that the dipole moments at each side are
oriented in opposite directions. The second chain consists also of
a sequence of identical base pairs but there exists a single
dipole moment which is oriented in opposite direction to that of
the others ones. I both cases there exists a local inhomogeneity
defined by the interface or the single inverted dipole,
respectively.

  The Hamiltonian of the Peyrard--Bishop model is augmented
with an energy term that takes into account the long--range forces
due to the dipole--dipole coupling between the base pairs. We have
generated MBs with different translational kinetic energies and
studied the effects of the collisions with the interface or with
the single inverted dipole, respectively. In each case, the
outcome depends on the values of the stacking coupling constant,
the dipole--dipole coupling constant, the frequency and the
translational kinetic energy of the MBs. Taking  fixed values for
the first three quantities, we have found that if the kinetic
energy of the MB is smaller than a critical value, the breather is
trapped after colliding with the interface, whereas if the kinetic
energy is larger, the breather is transmitted to the other side of
the interface loosing energy. Therefore, there exist two different
dynamical regime: a)~the trapping regime, when MBs are trapped at
the interface; b)~the transmission regime, when MBs are
transmitted to the other side of the interface.
  The same effects appear when MBs collide with a single
inverted dipole in the chain.
  In both cases, a necessary condition for the trapping of a MB is
  the existence of a stationary breather solution with the same frequency, centered at
the inhomogeneity site, which we call inhomogeneity breather (IB),
and with a smaller energy than the internal energy of the MB. When
collision occurs part of the energy is transferred to the chain as
phonon radiation. The existence, for each case, of a symmetric
effective on--site potential well due to the dipole-dipole
coupling, explains qualitatively the found behavior.

The value of the critical translational kinetic energy that
determines the transition from the trapping regime to the
transmission regime is directly related to the extent of the
localization of the IB. This extent of localization is related to
the quotient between the long--range and the short--range coupling
parameters. Trapping or accumulation of energy is facilitated when
this quotient increases.

 It is important to state that the most simple structural modification in a homogeneous DNA
 molecule, which is to reverse the orientation of a single dipole,
provides a mechanism for the trapping of moving breathers. The
simulations show that some incoming MBs can also be trapped after
colliding with a trapped breather, producing an accumulation of
energy at the inhomogeneity region. The interface or the inverted
dipole can play the role of chargers of vibrational energy.
Occasionally, a successful incoming breather can collide with this
excited region and all the accumulated energy can be carried away
to the rest of the DNA chain.
 At this point, if these processes
occur really in nature, one is tempted to imagine that succsesive
pulses of accumulated energy can arrive to a coding sequences
producing the initiation of the transcription processes. The
design of experiments in real or synthetic DNA are encouraged in
order to verify the results presented in this paper.

\section*{Acknowledgments}
We would like to thank Professor Yu B Gaididei and Professor JM
Romero for helpful discussions.

 This work has been supported by the European
Commission under the RTN project LOCNET, HPRN-CT-1999-00163 and
the MECD--FEDER project FIS2004-01183.

\newcommand{\noopsort}[1]{} \newcommand{\printfirst}[2]{#1}
  \newcommand{\singleletter}[1]{#1} \newcommand{\switchargs}[2]{#2#1}

\end{document}